\documentclass[3p, twocolumn]{elsarticle}
\usepackage{amssymb}
\usepackage{amsmath}
\usepackage{graphics}
\usepackage{graphicx}
\usepackage[hyperindex=true,final]{hyperref}

\begin{document}

\title{Aging Behaviour of the Localization Length in a Colloidal Glass}

\author[IPCF,DIP]{R.~Angelini\corref{cor1}}

\author[FEL]{A.~Madsen}

\author[BK]{A.~Fluerasu}

\author[DIP,IIT]{G.~Ruocco}

\author[IPCF,DIP]{B.~Ruzicka}

\cortext[cor1]{Corresponding author. Email:
roberta.angelini@roma1.infn.it, Tel: +390649913433, Fax:
+39064463158.}

\address[IPCF]{IPCF-CNR, I-00185 Roma, Italy}
\address[DIP]{Dipartimento di Fisica, Sapienza Universit$\grave{a}$ di Roma, I-00185, Italy}
\address[FEL]{European X-Ray Free Electron Laser, Albert-Einstein-Ring 19, D-22761 Hamburg, Germany}
\address[BK]{Brookhaven National Laboratory, NSLS-II, Upton, NY 11973, USA}
\address[IIT]{Center for Life Nano Science, IIT@Sapienza, Istituto Italiano di Tecnologia, Viale Regina Elena 291, 00161 Roma, Italy.}

\date{\today}

\begin{abstract}

The localization length $r_{loc}$ associated with a fast secondary
relaxation in glassy Laponite is determined by X-ray photon
correlation spectroscopy (XPCS) through a Debye-Waller fit of the
non-ergodicity parameter. Quantitative differences are observed
between the time dependence (aging) of $r_{loc}$ in spontaneously
aged and rejuvenated samples. This behavior is also reflected in
the calculated shear modulus which matches well with data obtained
by rheological measurements.
\end{abstract}

\maketitle


\vskip 10cm

\noindent\textbf{1. Introduction}

Colloidal suspensions are very versatile systems for the study of
aging dynamics towards arrested states. By changing parameters
like  ionic strength and  particle concentrations, they offer the
possibility of exploring for instance
gels~\cite{ZaccarelliJPCM2007, Lu_Nat_2008} and
glasses~\cite{PuseyNat1986, Imhof_PRL_1995} in a systematic way.
These systems are typically characterized by two relaxation
processes: a fast microscopic relaxation that is almost age
independent and a slower structural relaxation which changes with
age. The most direct way of accessing microscopic information
about the aging process is to study the time evolution of the
dynamic structure factor. This allows to extract the
characteristic times of the system (relaxation times) as well as
their distributions ($\beta_Q$ exponent - see description of the
fitting procedure further below). For this purpose, Dynamic Light
Scattering (DLS) and X-ray Photon Correlation Spectroscopy (XPCS)
are the most suitable techniques which enable to explore a large
range of time and length scales (set by the momentum transfer Q).

Here we report XPCS measurements on a  charged colloidal system,
Laponite. This colloidal clay is characterized by a complex phase
diagram with multiple arrested states depending on  clay and salt
concentrations. In salt free water, two types of arrested states,
an equilibrium gel~\cite{Ruzicka_NatMat_2011} and a Wigner
glass~\cite{Ruzicka_PRL_2010} can be distinguished at weight
concentration intervals 1.0 $\le C_w <$ 2.0 \% and 2.0 $\le C_w
\le$ 3.0 \%, respectively. Here, we focus on the glassy state at
$C_w=$3.0 \% where a very recent XPCS
study~\cite{AngeliniSoftMat2013} reported a dichotomic long time
aging behavior of the correlation functions: stretched ($\beta_Q
<1$) for spontaneously aged samples and compressed ($\beta_Q>1$)
for rejuvenated ones. Supposedly, shear applied by the syringe
induces internal stresses that are responsible for the compressed
behavior of the rejuvenated sample. In the aforementioned study
the fast relaxation time could not be accessed directly because it
falls outside the detection window accessible by XPCS. Therefore,
in order to address this point we perform in the present work  a
new specific analysis. We find that the fast dynamics display
different aging behaviors for the spontaneously aged and the
rejuvenated samples. Specifically, the aging dependence of the
Debye Waller factor and of the localization length $r_{loc}$ is
different in the two cases and mimics the difference between the
slow dynamics found earlier~\cite{AngeliniSoftMat2013}. The
difference is further elucidated by calculating the elastic shear
modulus $G_g'$ obtained through $r_{loc}$ and the static structure
factor by applying the standard Green-Kubo relation with a
factorization approximation~\cite{Gotze_RPP_1992, Gotze_JPCM_1999,
Schweizer_JCP_2007}.

\noindent\textbf{2. Materials and Methods}

\noindent\textit{2.1 Materials}

Laponite is a synthetic clay that, when dispersed in water, forms
a charged colloidal suspension of platelets with 25 nm diameter
and 0.9 nm thickness and inhomogeneous charge distribution,
negative on the faces and positive on the rims. The platelets are
usually considered monodisperse in size but a small polydispersity
has been reported by different
authors~\cite{Kroon_PRE_1996,Balnois_LANG_2003}.

Aqueous dispersions of Laponite RD with weight concentrations
C$_w=3.0 \%$  were prepared according to the protocol  described
in~\cite{Ruzicka_SoftMat_2011} ensuring reliable and reproducible
samples. The entire preparation process takes place in a glovebox
under $N_2$ flux to prevent CO$_2$
degradation~\cite{Thompson_JCIS_1992}. Laponite powder,
manufactured by Laporte Ltd., is dispersed in pure deionized
water, stirred vigorously for 30 min, and filtered soon thereafter
through a 0.45 $\mu$m pore size Millipore filter. A part of the
stock solution is directly filtered in glass capillaries of 2 mm
diameter for the experiments. These are later referred to as
``spontaneously aged" samples. The origin of the waiting time
($t_w=0$) determines the age of the sample and for the
spontaneously aged samples it is the time at which the suspension
is filtered. Rejuvenated samples are prepared from the stock
solution that had rested some time $t_R$ (rejuvenation time) since
filtration. The age $t_w$ of a rejuvenated sample is counted from
$t_R$ {\em i.e.} the time at which it was taken from the stock and
injected into the capillary by a syringe, hence introducing a huge
shear field (shear rejuvenation).

\noindent\textit{2.2 Measurements}

The samples were characterized by X-ray Photon Correlation Spectroscopy (XPCS) \cite{Madsen_NJP_2010,Leheny_COCIS_2012} at
beamline ID10A of the European Synchrotron Radiation Facility (ESRF) in Grenoble. For the measurements a partially coherent and
monochromatic X-ray beam with a photon energy of 8 keV was employed. Long series of scattering images were recorded by a
charged coupled device (CCD) placed in the forward scattering direction. The images were post processed following the
multi-speckle XPCS approach \cite{Madsen_NJP_2010} to get access to the dynamics of the samples. Ensemble averaged intensity
autocorrelation functions $g_2(Q,t)=\langle \frac{\langle I(Q,t_0)I(Q,t_0+t)\rangle_p}{\langle I(Q,t_0)\rangle_p \langle
I(Q,t_0+t)\rangle_p }\rangle_{t_0}$ were calculated using a standard multi-tau algorithm. Here, $\langle...\rangle_p$
indicates averaging over pixels of the detector mapping onto a single value of the momentum transfer ($Q$) while
$\langle...\rangle_{t_0}$ indicates temporal averaging over $t_0$.

\noindent\textbf{3. Results and Discussion}

\begin{figure}[t!]
\centering
\includegraphics[width=8cm,angle=0,clip]{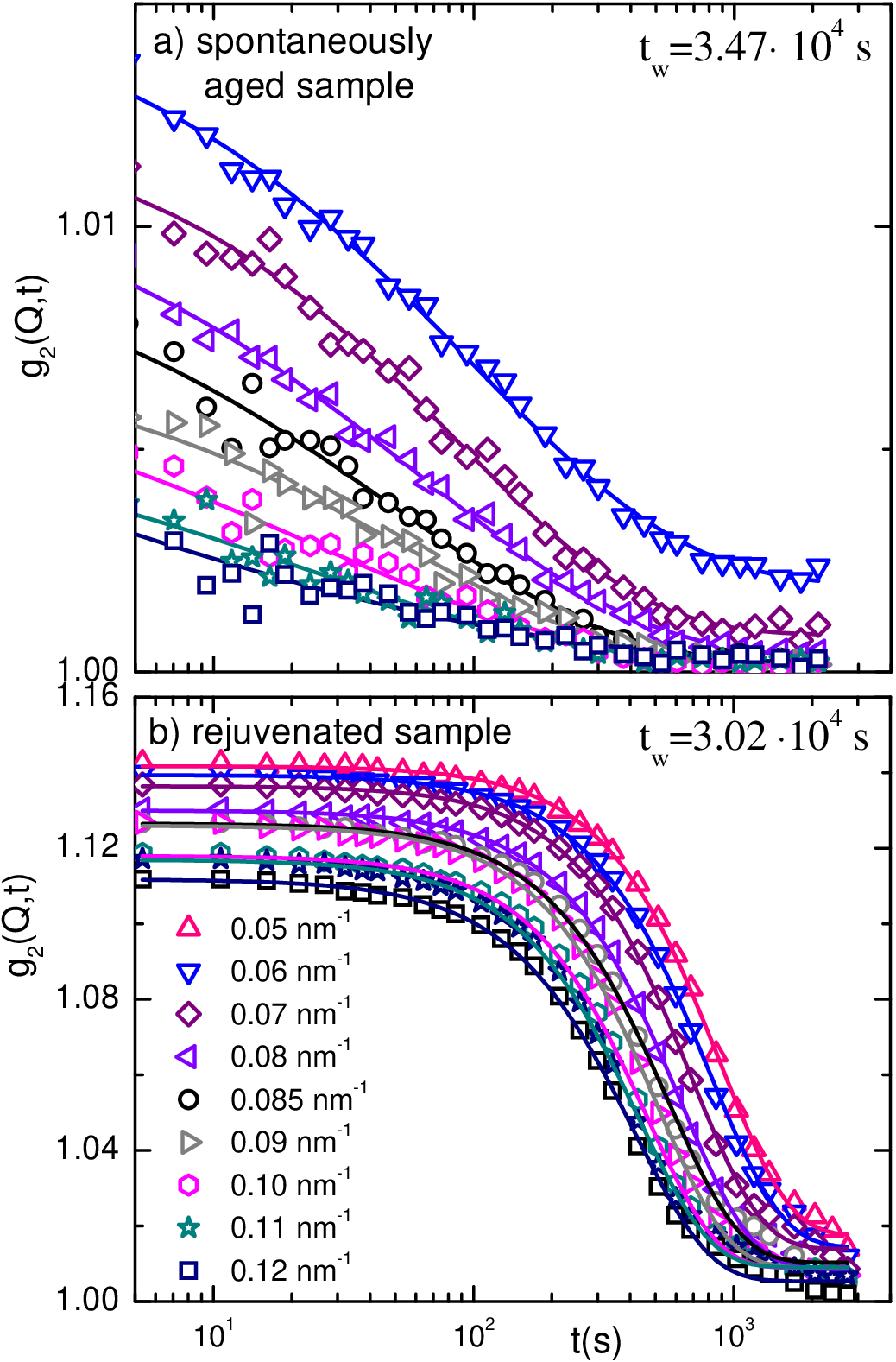}
\caption{Intensity autocorrelation functions of aqueous Laponite
suspensions (C$_w$=3.0 $\%$) at different $Q$ values (symbols) for
(a) a spontaneously aged sample at $t_w$=3.47 $\times 10^4$ s, and
for (b) a rejuvenated sample with $t_R$= 3.5 days and $t_w$=3.02
$\times 10^4$ s. The solid lines show the best fits performed by
Eq.~\ref{eq1}} \label{Fig1}
\end{figure}

Figure~\ref{Fig1} shows, as an example, the intensity autocorrelation  functions at different $Q$ values of Laponite
suspensions that are either spontaneously aged (Fig. ~\ref{Fig1}a) or rejuvenated (Fig. ~\ref{Fig1}b). For both samples, and at all
aging times and $Q$, the XPCS data are well described by the expression

\begin{equation}\label{eq1}
g_2(Q,t)-1 \propto (C\exp(-(t/\tau_Q)^{\beta_Q}))^2 = C^2|f(Q,t)|^2,
\end{equation}

where $f(Q,t)$ is known as the intermediate scattering function. $C^2$ represents the contrast, $\tau_Q$ the relaxation time and
$\beta_Q$ the Kohlrausch exponent. The latter two parameters characterize the dynamics of the sample. The fits with
Eq.~\ref{eq1} are shown as full lines in Fig.~\ref{Fig1}.

\begin{figure}[t!]
\centering
\includegraphics[width=8cm,angle=0,clip]{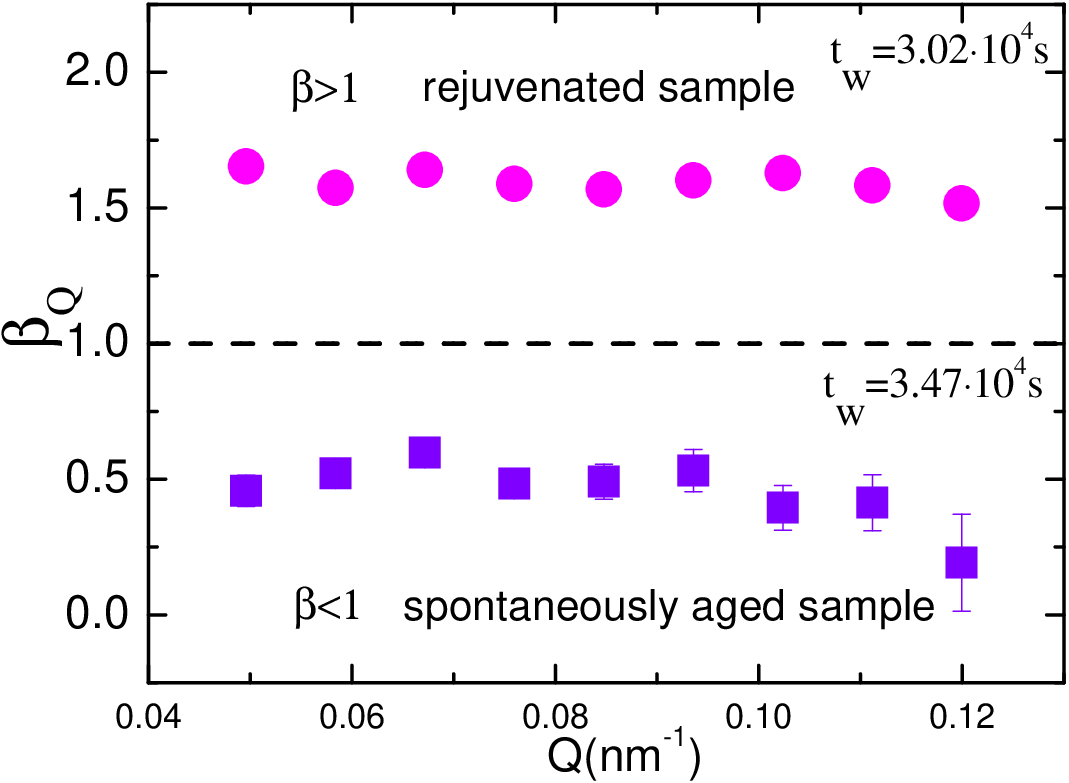}
\caption{Kohlrausch exponent $\beta_Q$ as a function of Q obtained
from Eq.~\ref{eq1} for the spontaneously aged ($\beta_Q<1$) and
the rejuvenated ($\beta_Q>1$) sample at C$_w$=3.0 $\%$.}
\label{Fig2}
\end{figure}

The exponent $\beta_Q$ obtained by the fit analysis is plotted vs
$Q$ in Fig.~\ref{Fig2} for both the spontaneously aged and the
rejuvenated sample. As recently reported in
Ref.~\cite{AngeliniSoftMat2013} the value of $\beta_{Q}$ is always
well below 1 for  the spontaneously aged sample in the $t_w$ and
$Q$ ranges investigated here. This means that the correlation
function in Eq.~\ref{eq1} takes a stretched exponential form which
is commonly observed for glass dynamics. For the rejuvenated
sample, $\beta_Q$ is always above unity implying that the decay of
the correlation function is faster than exponential, {\em i.e.} a
so-called compressed exponential behavior is observed. This has
often been observed in gel~\cite{Cipelletti_PRL_2000, GuoPRE2010}
and glassy systems~\cite{Bellour_PRE_2003, Bandyopadhyay_PRL_2004,
Schoesseler_PRE_2006}. Recently, compressed exponential decays of
$g_2(Q,t)$ have also been found in a metallic glass
\cite{Ruta_PRL_2012} and there appears to be a connection with the
development and relaxation of stresses in samples that can no
longer flow \cite{Cipelletti_FD_2003}.

\begin{figure}[t]
\centering
\includegraphics[width=8cm,angle=0,clip]{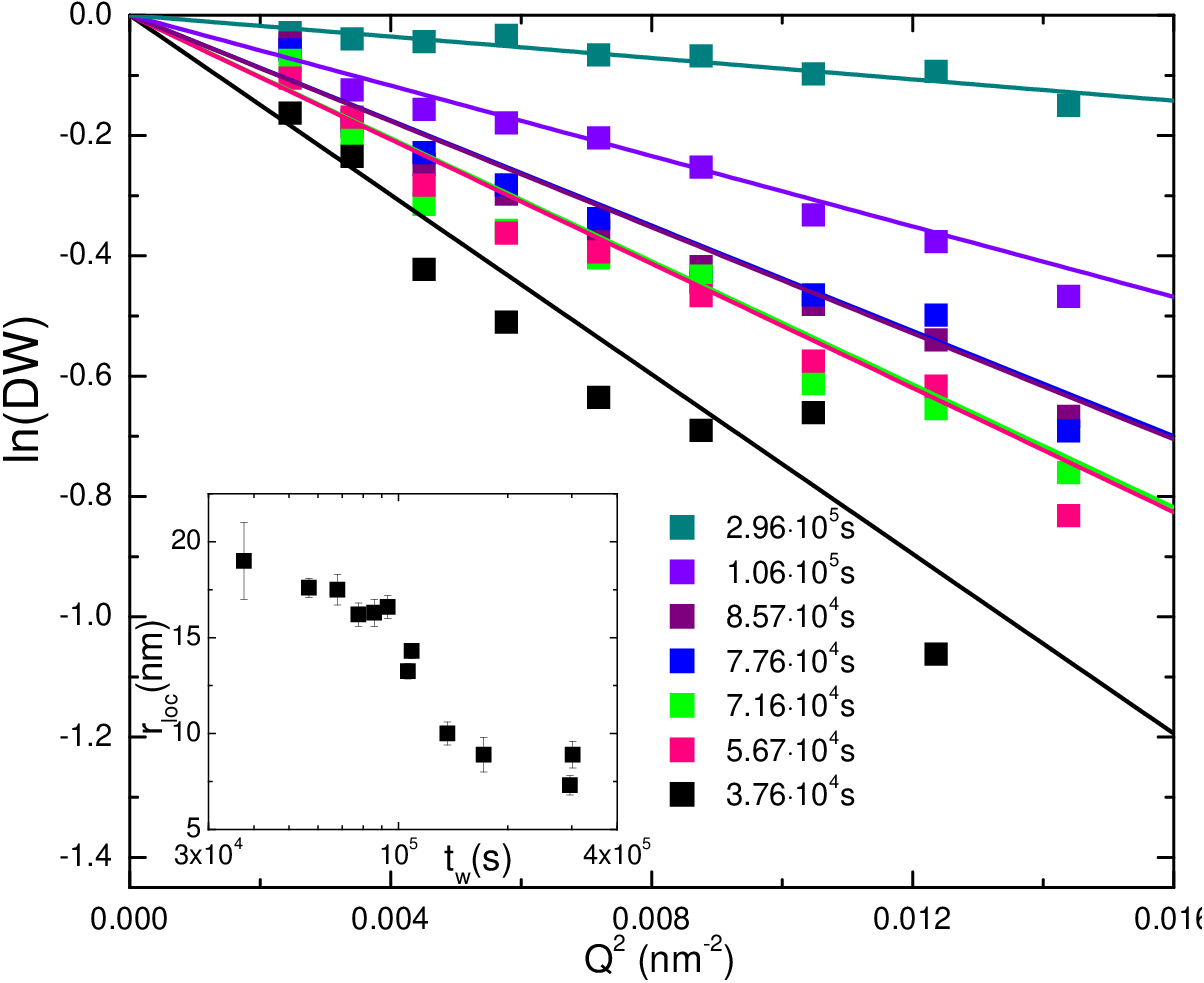}
\caption{$ln(C/C_0)$ data (symbols) and $ln(\text{DW})$  (lines)
as a function of $Q^2$ at different ages for spontaneously ages
sample with C$_w$=3.0$\%$. Inset: Age evolution of the
localization length $r_{loc}$} \label{Fig3}
\end{figure}

From multi-speckle XPCS measurements, the non-ergodicity parameter
can be derived.  It corresponds to the short time plateau value
($C$, Eq.~\ref{eq1}) of the intermediate scattering function
$f(Q,t)$~\cite{GuoPRE2010}. A value of $C$ smaller than $C_0$
(value expected from the optical contrast) indicates ``missing
contrast" and thus the presence of an additional decay of the
correlation function that is too fast to be measured directly. A
variation of $C$ with $Q$ is a strong signature of fast dynamics
in the sample and by inspecting Fig.~\ref{Fig1} it can be seen
that this is indeed the case here.

The Debye-Waller factor (DW) describes the reduction in the intermediate scattering function due to time averaged uncorrelated
dynamics, {\em e.g.} thermal motion. The DW equals $\langle \exp(i\mathbf{Q}\cdot\mathbf{u}(t)) \rangle$ where $\mathbf{u}(t)$
describes the displacement of the scatterer away from the average position. Expanding the DW to 2nd order terms in
$\mathbf{Q}\cdot\mathbf{u}$ and using the fact that $\langle \mathbf{Q}\cdot\mathbf{u} \rangle$=0 for random uncorrelated dynamics, one gets

\begin{equation}\label{eq2}
\text{DW} \approx 1-\frac{1}{6} Q^2 \langle u^2 \rangle \approx \exp(-(Q^2 r_{loc}^2)/6)
\end{equation}

where the localization length $r_{loc}=\sqrt{\langle u^2
\rangle}$.  The reduced contrast $C/C_0$ in Fig.~\ref{Fig1} can be
well described by the DW form (Eq.~\ref{eq2}) and permits to
derive the localization length $r_{loc}$ that characterizes the
range of localized colloidal motion in the Laponite suspensions.
In Fig.~\ref{Fig3} the quantity $ln(\text{DW})$ is plotted vs
$Q^2$  for a spontaneous aged sample at different waiting times
(ages) and compared with the data. The inset of the figure shows
the age behaviour of $r_{loc}$ obtained from fits with
Eq.~\ref{eq2} to the data. Both the $ln(\text{DW})$ and the
$r_{loc}$ behaviour found for the spontaneously aged sample are in
qualitative agreement with the results reported  by Bandyopadhyay
et al. in~\cite{Bandyopadhyay_PRL_2004} where
$<\bar{r}^2>^{1/2}=r_{loc}/\sqrt{2}$. In that case however, and at
variance with our results, the intensity autocorrelation functions
show compressed behaviours. A clear understanding of differences
and similarities with respect to other published data is not
always possible but it is suggested that variations in preparation
protocol can be responsible for the discrepancies.

\begin{figure}[t]
\centering
\includegraphics[width=8cm,angle=0,clip]{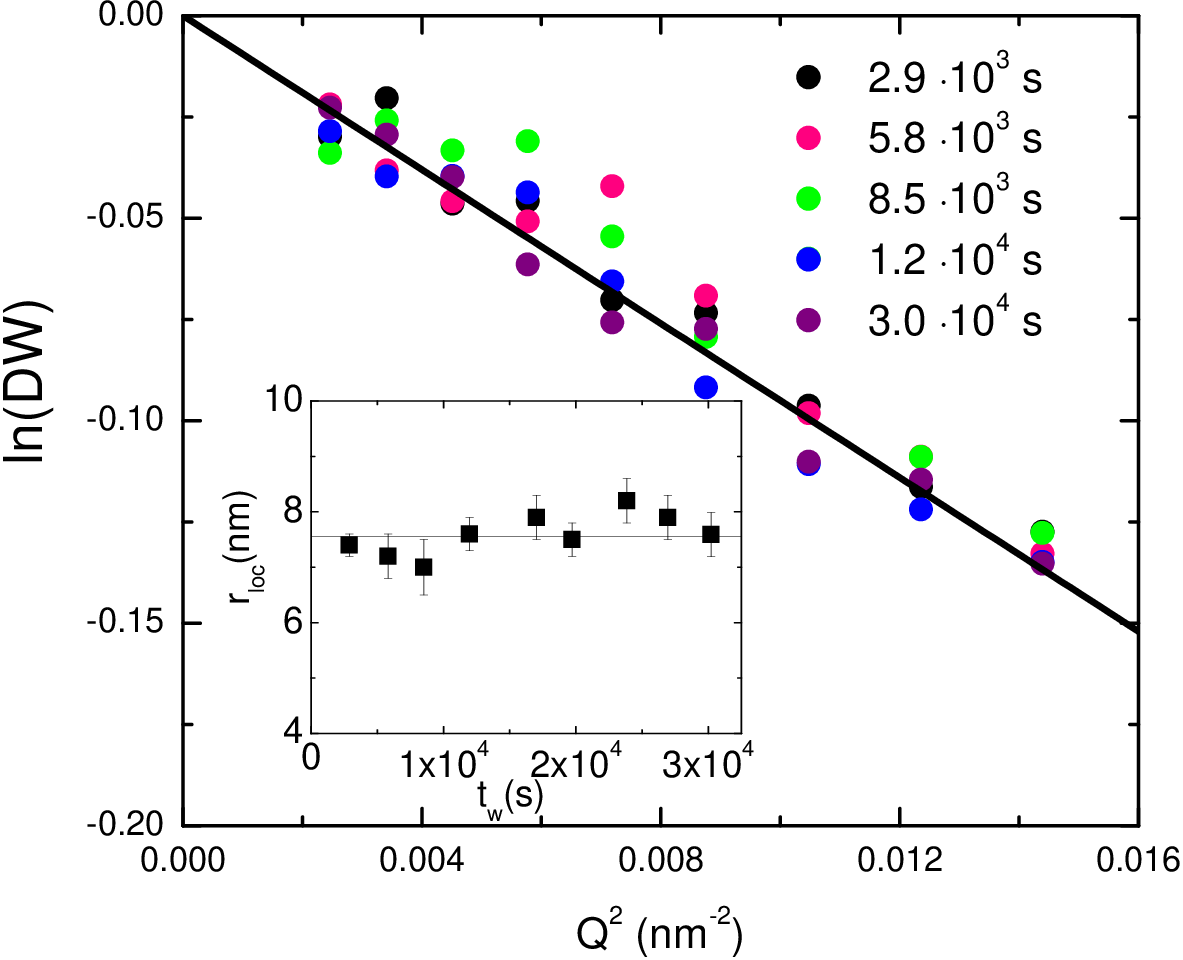}
\caption{$ln(C/C_0)$ data (symbols) and $ln(\text{DW})$ (lines) as
a  function of $Q^2$ at different ages for a rejuvenated sample
with C$_w$=3.0$\%$ and $t_R$=3.5 days. Inset: Age evolution of the
localization length $r_{loc}$} \label{Fig4}
\end{figure}

In Fig.~\ref{Fig4} the same analysis is shown on a rejuvenated
sample ($t_R=$3.5 days) from the same batch and hence of the same
concentration. In the investigated time window, the localization
length is smaller than for the spontaneously aged sample (inset of
Fig.~\ref{Fig3}) which indicates a harder structure with slower
structural relaxations for the rejuvenated suspension. This is
confirmed by a detailed study of the structural relaxation times
$\tau_Q$ \cite{Ianni_PRE_2007}, which in particular shows that the
shear rejuvenation is not fully rewinding the dynamics {\em i.e.}
it remains slower than for a spontaneously aged sample of same
$t_w$.

The spontaneously aged suspension displays a linear decrease of
$r_{loc}$  with age (inset of  Fig.~\ref{Fig3}). Hence, the
stiffening of the structure on the local scale is consistent with
the increase of $\tau_Q$ with age that has been observed earlier
\cite{AngeliniSoftMat2013}. On the contrary, for the rejuvenated
sample $r_{loc}$ remains constant with age (inset of
Fig.~\ref{Fig4}). This difference is another indication of the
fundamental change in microscopic dynamics that happens upon
rejuvenation. Hitherto, this phenomenon was only qualitatively
described as a change from stretched ($\beta_Q<1$) to compressed
($\beta_Q>1$) exponential decay of the intensity correlation
functions but now, in addition, it manifests itself by the age
dependence and magnitude of $r_{loc}$.

To further explore these new results we have calculated the storage modulus $G'_g$ from $r_{loc}$ as suggested by Schweizer and
coworkers~\cite{Gotze_RPP_1992, Gotze_JPCM_1999,Schweizer_JCP_2007}:

\begin{equation}
G'_{g}=\frac{k_BT}{60
\pi^2}\int_0^{\infty}dk[k^2\frac{d}{dk}ln(S(k))]^2\exp\{-k^2r^2_{loc}/3S(k)\},
\label{eq3}
\end{equation}
where $S(k)$ is the static structure factor. $S(k)$ has been
measured for both spontaneously aged and rejuvenated samples and
it has been used to calculate the integral of Eq.\ref{eq3} with
the approximation of using real Q values as integration limits.

\begin{figure}[t]
\centering
\includegraphics[width=7cm,angle=0,clip]{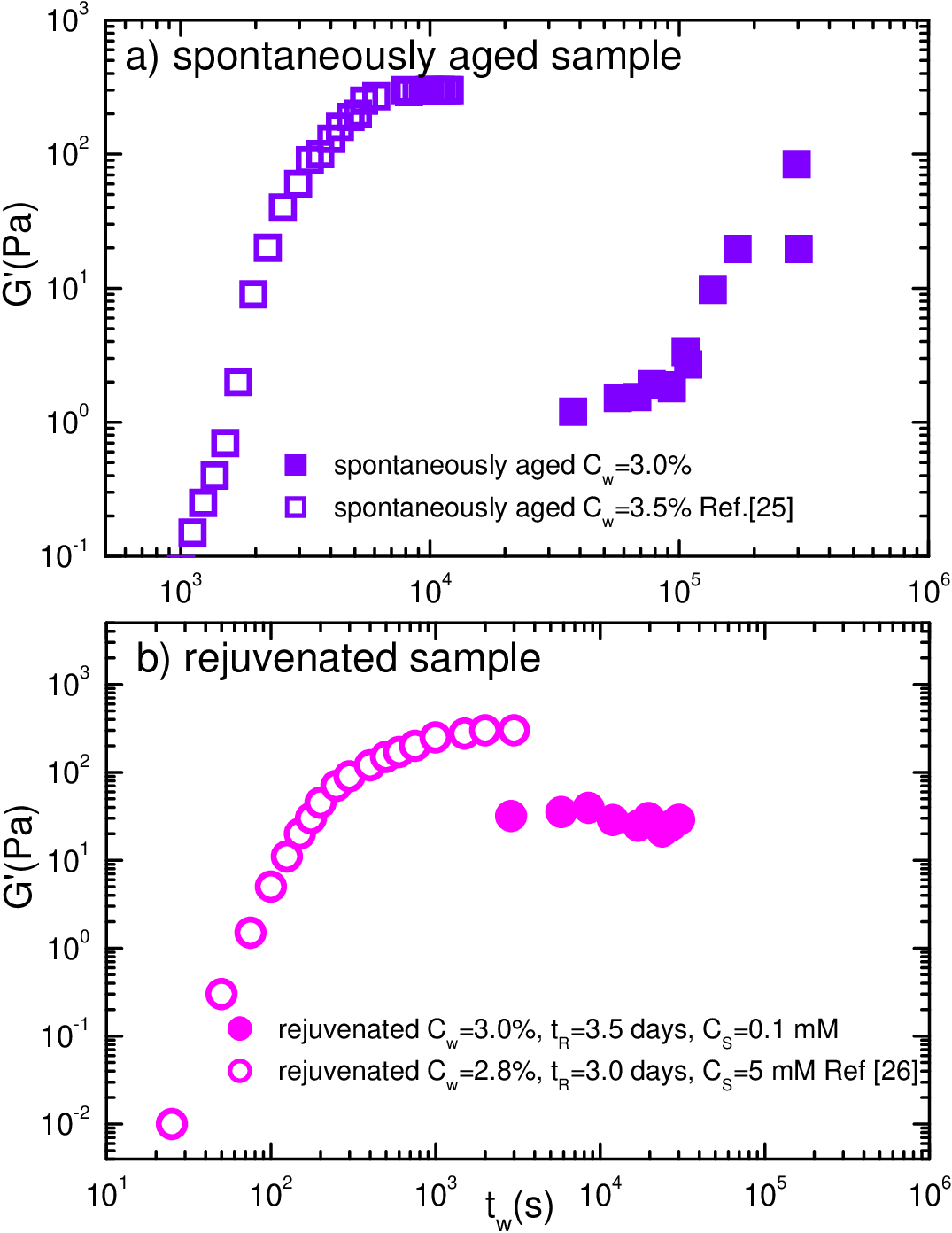}
\caption{Time evolution of the storage modulus $G'$ as derived
from $r_{loc}$ (see text) for Laponite suspensions with $C_w$=3.0
\% in salt free water ($C_s=10^{-4}$ M) (full symbols). (a)
Spontaneously aged sample (full squares) compared with rheological
measurements on a $C_w$=3.5 \% sample of the same salt
concentration by Ref.~\cite{Bonn_EL_1998} (open squares). (b)
Rejuvenated sample at $t_R$=3.5 days (full circles) compared with
rheological measurements on a $C_w$=2.8 \%, $C_s$=5 mM at
$t_R$=3.0 days sample by Ref.~\cite{Shahin_Langmuir_2010} (open
circles)} \label{Fig5}
\end{figure}

Fig.~\ref{Fig5} shows $G'_g$ calculated from Eq.~\ref{eq3} as a
function of waiting time for a spontaneously aged
(Fig.~\ref{Fig5}a, full symbols) and a rejuvenated sample
(Fig.~\ref{Fig5}b, full symbols). In the case of a spontaneously
aged sample at $C_w$=3.0 \% (Fig.~\ref{Fig5}a) an increase of
$G'_g$ (full squares) by almost two orders of magnitude is
observed with age, indicating that the sample is becoming more and
more solid-like. This behaviour is similar to that obtained by
rheological measurements on a $C_w$= 3.5 \% sample (empty
squares)~\cite{Bonn_EL_1998} shown in the figure. The difference
between the two curves could be due to different clay
concentrations and/or the approximations in the model calculation
of $G'_g$.

The rejuvenated sample (Fig.~\ref{Fig5}b) shows a calculated
$G'_g$ (full circles) that is almost constant with waiting time
and $G'_g \sim$ 30 Pa consistent with a solid-like nature. This
behaviour is compared with that obtained by rheological
measurements on a rejuvenated $C_w$= 2.8 \% sample at higher salt
concentration ($C_s$=5 mM) and $t_R$=3 days (empty
circles)~\cite{Shahin_Langmuir_2010} also shown in figure
\ref{Fig5}b. In this case the higher plateau value of the
rheological measurement can possibly be attributed to a higher
salt concentration~\cite{Shahin_Langmuir_2010}.

\noindent\textbf{4. Conclusions}

In conclusion, we have quantified the difference between  a
spontaneously aged and a rejuvenated Wigner glass (Laponite
suspension) by studying the Debye-Waller factor of the
intermediate scattering functions obtained by XPCS. The
Debye-Waller factor allows deriving $r_{loc}$, the localization
length of the fast dynamics of the system. While $r_{loc}$ remains
constant for the rejuvenated suspension it decreases steadily with
age for the spontaneously aged sample. From the present data it is
not possible to conclude whether $r_{loc}$ stabilizes with long
enough waiting times. The observed behaviours and comparison with
previous results~\cite{Bandyopadhyay_PRL_2004} indicate that
Laponite suspensions are not only sensitive to sample preparation
and solution conditions but also dependent, in the case of
rejuvenated samples, on other conditions as applied shear field,
shear history, idle time before application of
shear~\cite{Shahin_Langmuir_2010}, etc. Evidently, this
complicates comparison of results obtained by different
preparation protocols and methods. However, in the present case
the calculated shear modulus $G'_g$ is found to be compatible with
earlier rheological measurements.

For both samples the fast dynamics display a dichotomic aging
behaviour, rather similar to the observations of the slow
dynamics~\cite{AngeliniSoftMat2013} relaxations. While for
spontaneously aged samples we find stretched correlation functions
(slow relaxation) and a correlation length (fast relaxation) that
decreases with waiting time, for rejuvenated samples compressed
behaviour (slow relaxation) and a time constant correlation length
(fast relaxation) is observed. This indicates a coupling between
the fast and slow dynamics in the two systems.

\noindent\textbf{Acknowledgments}

We acknowledge ESRF for beamtime for this project. The staff at
beamline ID10A is acknowledged for help during the experiments. AF
acknowledges support from DOE grant EAC02-98CH10886 (NSLS-II
project).

\end{document}